\documentclass[a4paper,11pt]{article}
\usepackage{jcappub} 
\usepackage{lineno}

\usepackage{url}
\usepackage{wasysym} 
\usepackage{xcolor}
\usepackage{makecell}
\usepackage{graphicx}
\usepackage{multirow}

\newcommand{\sv}{\ensuremath{\langle\sigma v\rangle}}

\newcommand{\cms}{\ensuremath{\rm cm^3~s^{-1}}}

\title{Constraining the dark matter origin of the halo-like 20 GeV $\gamma$-ray excess with the AMS-02 antiproton data}

\author{Xiao Wang$^{a,b}$,}
\author{Kai-Kai Duan*$^a$}
\affiliation{$^a$Key Laboratory of Dark Matter and Space Astronomy, 
Purple Mountain Observatory, Chinese Academy of Sciences, Nanjing 210033, China}
\affiliation{$^b$School of Astronomy and Space Science, University of Science and Technology of China, Hefei, Anhui 230026, China}

\emailAdd{duankk@pmo.ac.cn}

\abstract{Very recently,  a significant $\sim 20$ GeV gamma-ray excess in the Milky Way halo has been reported and  
a dark matter origin has been suggested. The inferred dark matter parameters are 
\( m_\chi \sim 0.5-0.8 \) TeV and \( \langle \sigma v \rangle \sim (5-8) \times 10^{-25} \) \cms for the \( b\bar{b} \) channel. If correct, prominent antiproton emission is produced and can be directly tested by the AMS-02 data. In this work we calculate the corresponding antiproton emission and show that the expected flux at $\sim 100$ GeV is already above the AMS-02 observation. A proper treatment on the antiproton background resulting from the high energy cosmic ray propagation would suggest an annihilation cross section of $< 2\times 10^{-26}$ \cms, which is a few$\times 10$ times lower than that needed to interpret the potential signal. We therefore conclude that the  $\sim 20$ GeV gamma-ray excess in the Milky Way halo is not a viable dark matter signal.
}

\begin{document}

\maketitle

\flushbottom

\newpage

\section{Introduction}
The astronomical observations have found compelling evidence for the presence of a large amount of dark matter (DM) in the universe. The nature of the dark matter however remains unclear.  So far, various dark matter particle candidates have been suggested in the literature. Among them, the weakly interacting massive particles (WIMPs) have attracted wide attention and dedicated efforts have been made to hunt them~\citep{2018RvMP...90d5002B, 2021PrPNP.11903865A}. One promising approach is to identify the dark matter signal in cosmic rays and gamma-rays supposing the dark matter can annihilate or decay into the pairs of the normal particles. Some intriguing features in cosmic ray electrons, positrons, antiprotons, as well as gamma-rays had been reported previously~\citep{2008Natur.456..362C,2009Natur.458..607A,2009arXiv0910.2998G, 2015PhRvD..91l3010Z, 2017ApJ...840...43A,2017PhRvL.118s1101C, 2017PhRvL.118s1102C,2017Natur.552...63D,2024arXiv240711737F}, but their dark matter origin can not be established. One good example is the so-called  Galactic Center Excess (GCE), a distinct GeV gamma-ray excess in the direction of the Galactic center with a spectrum and spatial distribution that can be well reproduced by the annihilation of the $\sim 40-60$ GeV dark matter particles~\citep{2009arXiv0910.2998G, 2015PhRvD..91l3010Z, 2017ApJ...840...43A}. However, because of the very complicated background emission in the direction of the Galactic center, it is unclear whether this excess is actually from a group of unresolved millisecond pulsars or even the stellar bulge~\citep{2020ARNPS..70..455M}.   

Nevertheless, the Galaxy constitutes a prime target for DM  detection with gamma-rays due to its proximity and relatively high dark matter density, which maximizes the expected signal from particle annihilation or decay. Its close distance ($\sim 8$ kpc) allows us to observe faint gamma-ray signals with high sensitivity, while the dense dark matter halo enhances the probability of interactions in particular the annihilation. 
Intriguingly, a recent analysis of 15 years of Fermi-LAT data reveals a statistically significant 20 GeV gamma-ray excess in the Milky Way halo~\citep{2025JCAP...11..080T}.
Its spectral and morphological properties are consistent with dark matter annihilation via an NFW-like profile, suggesting parameters $m_\chi\sim 0.5-0.8$ TeV and $\sv \sim (5-8)\times 10^{-25}~\text{cm}^3~\text{s}^{-1}$ for $b\bar{b}$ channel (for the $W^{+}W^{-}$ channel, similar parameters are needed).
However, this cross-section considerably exceeds standard thermal relic predictions~\citep{2018RvMP...90d5002B, 2021PrPNP.11903865A} and dwarf galaxy constraints \citep{2015PhRvL.115w1301A, 2017ApJ...834..110A} by approximately one order of magnitude.
To resolve this tension, Murayama~\citep{2025arXiv251201404M} proposed resonant dark matter annihilation as a solution to reconcile the conflicting cross-sections from the Milky Way gamma-ray signal, freeze-out requirements, and dwarf galaxy limits, conducting both model-independent parameter determinations and presenting a concrete particle physics realization.

In this work we take the AMS-02 antiproton data to directly constrain the dark matter annihilation origin of the 20 GeV gamma-ray excess reported in~\citep{2025JCAP...11..080T}. In both the $b\bar{b}$ and $W^{+}W^{-}$ channels, simultaneously with the generation of the gamma-rays, abundant antiprotons are produced as well. 
Hence the high quality antiproton data collected by AMS-02 can 
serve as a crucial consistency check~\citep{2016PhRvL.117i1103A, 2021PhR...894....1A, 2025PhRvL.134e1002A}. In line with previous tight constraints from AMS-02 data~\citep{2017PhRvL.118s1101C, 2017PhRvL.118s1102C, 2022PhRvL.129w1101Z, 2025JCAP...10..049D}, we demonstrate that the proposed cross-section by~\citep{2025JCAP...11..080T} would significantly exceed the measured antiproton flux limits. 


\section{The antiproton production and propagation}\label{sec: method}

\subsection{Propagation of CRs}
We employ the propagation model that combines the effects of diffusion and re-acceleration of cosmic rays (CRs) within the Milky Way~\citep{2007ARNPS..57..285S}.
The propagation equation can be represented as
\begin{align}
\label{eq:propagation}
    \frac{\partial \psi(\vec{r},p,t)}{\partial t}=& 
    \underbrace{q(\vec{r},p,t)}_{\rm source~term}+
    \underbrace{\nabla\cdot(D_{xx}\nabla\psi-\vec{V}\psi)}_{\rm diffusion~term}\nonumber\\
    &+
    \underbrace{\frac{\partial}{\partial p}p^2D_{pp} \frac{\partial}{\partial p} \frac{1}{p^2}\psi}_{\rm re-acceleration~term}- 
    \underbrace{\frac{\partial}{\partial p}\left[ \dot{p}\psi-\frac{p}{3}(\nabla\cdot\vec{V})\psi \right]}_{\rm energy~loss~term}-
    \frac{1}{\tau_f}\psi-\frac{1}{\tau_r}\psi,
\end{align}
where $\psi(\vec{r},p,t)$ is the phase space density distribution of CRs as a function of position $\vec{r}$, momentum $p$ and time $t$. 
The first term is the source of CR, $q(\vec{r},p,t)$, while the second and third terms are the diffusion and re-acceleration process, respectively. 
Based on the studies~\citep{2017PhRvD..95h3007Y, 2020JCAP...11..027Y}, 
the diffusion with re-acceleration scenario performs significantly better than the diffusion with convection scenario. 
Therefore, we omit the convection effect ($\vec{V} = 0$).
The fourth term includes energy losses and adiabatic losses. 
The terms $\tau_f$ and $\tau_r$ represent the fragmentation and radioactive decay time, respectively. 
Further details of the model can be found in \citep{2025JCAP...10..049D}.
The \texttt{GALPROP}~\citep{1998ApJ...509..212S} program\footnote{\url{https://galprop.stanford.edu}} is used to solve propagation equations in the interstellar medium (ISM) and calculates the local interstellar spectra (LIS) of CRs before they enter the solar system. Different CR particles are described by distinct but coupled propagation equations.

When CRs enter the solar system, their spectra are modulated by the solar wind and heliospheric magnetic field, especially for those with rigidity below 40 GV.
We will employ a simple force-field approximation (FFA)~\citep{1968ApJ...154.1011G} to model this physical process.
The FFA model provides an important analytical tool to calculate the spectra at the top of the atmosphere (TOA) using CR intensity measured in the ISM,
\begin{equation}
    J^{\rm TOA}(E) = J^{\rm LIS}(E + \Phi) \times \frac{E (E + 2 m)}{(E + \Phi)(E + \Phi + 2m)}\;,
    \label{equation:FFA}
\end{equation}
where $J^{\rm TOA}$ and $J^{\rm LIS}$ denote the cosmic-ray energy spectra at TOA and in ISM, respectively.
Here, $E$ is the particle kinetic energy per nucleon, $m$ is mass of proton, and $\Phi$ the force-field energy loss, given by $\Phi = Ze\phi / A$. 
In this expression, $Z$ and $A$ are the particle charge and mass number, $e$ is the elementary charge, and $\phi$ is the modulation potential, which typically ranges from 0.1 to 1 GV. 
For a detailed derivation of the FFA, see~\citep{1968ApJ...154.1011G}.

Since antiprotons are negatively charged, their solar modulation potential differs from that of positively charged particles such as protons~\citep{2016PhRvD..93d3016C,2022JCAP...10..051C}. Therefore, for each antiproton LIS, we independently determine the modulation potential \(\phi\).

\subsection{Antiprotons produced in the Galaxy}
CRs interacting with atoms in ISM through high-energy collisions generate secondary particles including antiprotons and $\gamma$-rays. This process applies to both cosmic-ray protons ($H$) and heavier nuclei ($N$), resulting in numerous secondary particles including daughter nuclei with atomic numbers lower than those of the initial CRs or ISM gas particles.
Within the \texttt{GALPROP} framework, the following nuclear interaction channels are considered for secondary particle production calculations: $H$-$H$, $H$-$N$, $N$-$H$, and $N$-$N$. The corresponding cross sections for these processes are adopted from~\citep{1983JPhG....9..227T}.
The production and propagation mechanisms of antiprotons are implemented following the prescriptions detailed in~\citep{2002ApJ...565..280M} and~\citep{2015ApJ...803...54K}.

An additional source of antiprotons arises from DM annihilation processes. 
We model the DM halo distribution according to the Navarro-Frenk-White (NFW) profile~\citep{1997ApJ...490..493N}, $\rho(r) = \rho_s[(r/r_s)(1+r/r_s)^2]^{-1}$, with $r_s=20~\rm{kpc}$ and $\rho_s=0.35~\rm{GeV}~\rm{cm}^{-3}$, centered on the Galactic center.
 
The source term \( Q(E, r) \) induced by dark matter annihilation is given by:
\begin{equation}\label{eq: DM-source}
    Q(E, r) = \frac{1}{2} \left( \frac{\rho(r)}{m_\chi} \right)^2 \langle \sigma v \rangle_{\text{ann}} \frac{\mathrm{d}N}{\mathrm{d}E},
\end{equation}
where \( E \) is the antiproton energy, and \( \frac{\mathrm{d}N}{\mathrm{d}E} \) is the antiproton energy spectrum produced per annihilation (averaged over annihilation events), which is adopted from~\citep{2011JCAP...03..051C, Ciafaloni2011Mar}.

\subsection{Uncertainty from propagation}
To avoid biases induced by uncertainties in Galactic propagation, we first tightly constrain the relevant transport parameters by minimizing $\chi^2 = \sum_i \left( \frac{{\rm Data}_i - {\rm Model}(E_i)}{\sigma_i} \right)^2$ using cosmic-ray data (protons, helium, C, O, B/C, B/O, and $^{10}$Be/$^{9}$Be) from AMS-02, DAMPE ACE-CRIS, and so on.
We combine the \texttt{GALPROP} code with the \texttt{emcee}~\citep{2013PASP..125..306F} package to perform a Markov Chain Monte Carlo (MCMC) analysis and obtain the posterior distributions of the key propagation parameters. Once the chains have converged, we extract a representative ensemble of parameter sets that faithfully sample the posterior distribution.

In the previous section, we discussed the two origins of cosmic-ray antiprotons. While antiprotons produced by dark matter annihilation depend solely on the source parameters and propagation parameters, secondary antiprotons are additionally sensitive to the energy spectra of primary cosmic rays (primarily protons and helium nuclei). To ensure accurate computation of the secondary antiproton yield, we set the maximum nuclear charge of cosmic-ray nuclei to $Z=14$.

With a sufficient ensemble of propagation samples, we compute the LIS for both antiproton components.
For the dark matter contribution, we directly adopt the parameters reported by Totani:
\begin{itemize}
    \item \( b\bar{b} \) channel: \( m_\chi = 510 \) GeV, \( \langle \sigma v \rangle = 6.3 \times 10^{-25} \) cm$^3$/s ;
    \item \( W^+W^- \) channel: \( m_\chi = 420 \) GeV, \( \langle \sigma v \rangle = 7.2 \times 10^{-25} \) cm$^3$/s .
\end{itemize}
Using these two parameter sets, we predict the corresponding antiproton spectra from dark matter annihilation.

As will be shown later, an excess antiprotons from dark matter would yield an unphysically large solar modulation potential if the total (background + DM) antiproton spectrum were used to fit \(\phi\). Moreover, since solar modulation primarily affects low energies while the present work focuses on the high-energy range, we fit the modulation potential solely to the background antiproton LIS. This procedure yields a physically reasonable value of \(\phi = 0.531 \pm 0.028\) GV.

\section{Constraint on the dark matter origin of the \texorpdfstring{$\sim$}{~} 20 GeV gamma-ray signal in the Galactic halo}\label{sec: results}

\begin{figure}
    \centering
    \includegraphics[width=\linewidth]{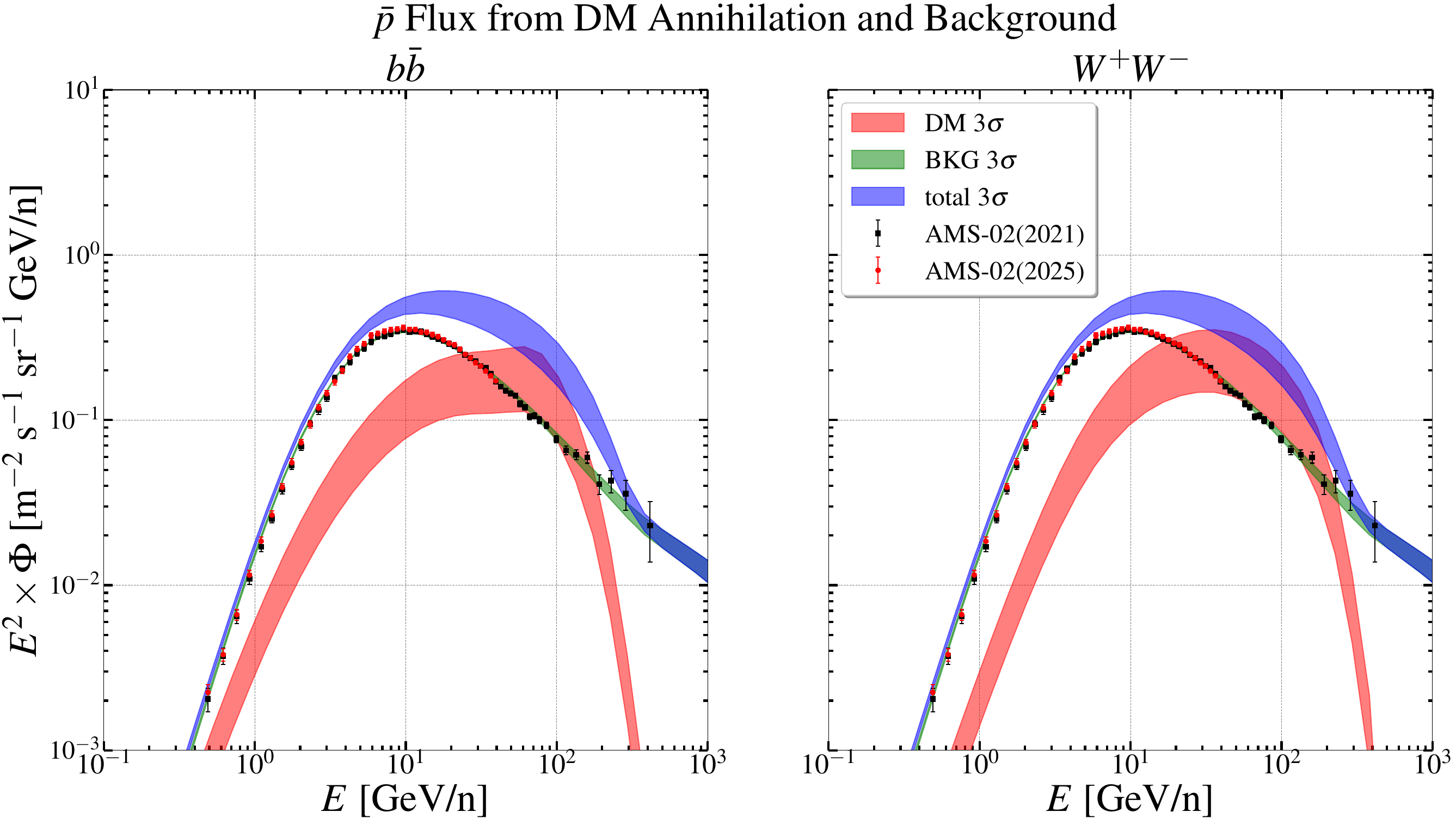}
    \caption{
    Antiproton spectra produced by dark matter annihilation through different channels. The left panel shows annihilation into $b\bar{b}$, while the right panel shows annihilation into $W^+W^-$. The green band represents the background antiproton flux (secondary production), the red band denotes the contribution from dark matter annihilation, and the blue band indicates the total (background + DM) flux. The error bars correspond to AMS-02 antiproton data published in 2021 (black) and 2025 (red). The total flux at 100 GeV exceeds the AMS-02 measurements by $15\sigma$ for the \( b\bar{b} \) channel and by $18\sigma$ for the \( W^+W^- \) channel. Here $\sigma$ refers to the experimental uncertainty of AMS-02 data.
    }
    \label{fig:excess-pbar}
\end{figure}

Using antiproton spectra computed with different propagation parameters, we determine the 0.15\% and 99.85\% quantiles in each energy bin to construct the corresponding 3$\sigma$ confidence band, as shown in Figure~\ref{fig:excess-pbar}. 
In both panels, three colored bands are displayed: the red band represents the antiproton spectrum produced solely by dark matter annihilation, the green band denotes the secondary antiprotons, and the blue band corresponds to the total spectrum obtained by summing the two contributions.  

It is readily apparent that, owing to the relatively large dark matter mass, the peak of the annihilation-induced spectrum shifts toward higher energies, misaligning with the peaks observed in the data and the background. Particularly around 100 GeV, the dark matter contribution alone reaches or exceeds the AMS-02 measured antiproton flux. This indicates that, even disregarding the secondary background, the antiprotons from dark matter annihilation are already strongly in 
tension with the experimental observations.

When the secondary antiprotons are included, the total flux substantially exceeds the data across the entire energy range considered, with deviations reaching beyond 20$\sigma$ in certain regions. This further underscores the incompatibility of the adopted dark matter parameters.


\begin{figure}
    \centering
    \includegraphics[width=0.8\linewidth]{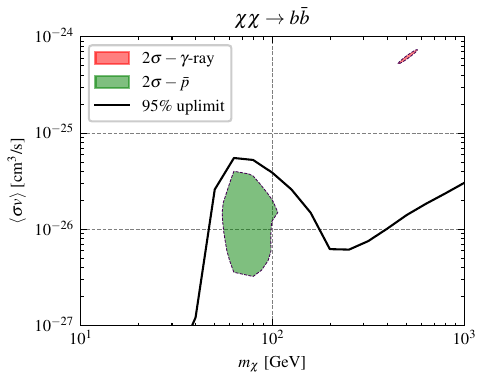}
    \caption{
    Comparison of dark matter annihilation signals derived from different astrophysical messengers. The red region in the upper-right corner represents the $2\sigma$ signal range inferred from Totani's work. The central green region corresponds to the $2\sigma$ annihilation signal obtained using AMS-02 antiproton data, while the black curve denotes the corresponding 95\% confidence-level upper limit. }
    \label{fig:contour}
\end{figure}

To mitigate biases arising from considering only the best-fit point, we seek to compare the signal intervals derived from both analyses. However, Totani~\citep{2025JCAP...10..049D} does not explicitly provide a dark matter signal interval, as systematic uncertainties dominate over statistical errors.
For completeness, we utilize the data points from the upper-left panel of his Figure 16 to compare with the DM-induced $\gamma$-ray spectrum (considering annihilation to \( b\bar{b} \) as an illustrative case).
Notably, the error bars in his Figure 16 are statistical 1$\sigma$ estimated by MCMC.
This yields a plausible \( 2\sigma \) signal interval, depicted as the red region in Figure~\ref{fig:contour}. Concurrently, we overlay the dark matter signal constraints derived from antiproton data in~\citep{2025JCAP...10..049D} (green region) along with the corresponding cross-section upper limit (black curve).

Evidently, Totani's signal region~\citep{2025JCAP...10..049D} exhibits substantial deviation from ours, exceeding our upper limit by nearly two orders of magnitude—a discrepancy that cannot be attributed to systematic errors. 
Integrating constraints from various observations on DM, including the WIMP thermal relic cross section and limits from dwarf spheroidal galaxies, we conclude that the dark matter signals and constraints derived from antiproton data are reasonably consistent with established benchmarks. In contrast, the parameters inferred from the 20 GeV gamma-ray excess deviate significantly from these results. Consequently, this excess is unlikely to originate from dark matter annihilation and may instead arise from other, as-yet-unidentified astrophysical sources.

\section{Discussion}\label{sec: conclusion}
One of the main purposes of the dark matter indirection detection is to identify the dark matter signal in the cosmic rays and/or gamma-rays. 
Dedicated efforts have been made to search for dark matter signal in the Fermi-LAT data. Though some interesting features have been reported in the literature, their dark matter origins are still in debate. This is the case even for the bright Galactic center excess peaking at a few GeV. 

Very recently, a new  analysis of 15 years of Fermi-LAT data reports a statistically significant 20 GeV gamma-ray excess in the Milky Way halo~\citep{2025JCAP...11..080T}.
Its spectral and morphological properties are consistent with dark matter annihilation via an NFW-like profile, but the inferred dark matter mass and annihilation rate are quite high.  Though  
a resonant dark matter annihilation model has been developed to reconcile the conflicting cross-sections from the Milky Way gamma-ray signal, freeze-out requirements, and dwarf galaxy limits~\citep{2025arXiv251201404M}, the antiproton data can directly constrain the dark matter origin of this $\sim 20$ GeV gamma-ray signal since both $b\bar{b}$ and $W^{+}W^{-}$ channels produce antiprotons simultaneous with gamma-rays. 

To achieve such a goal, we have calculated the production and the propagation of the antiprotons in the Galaxy. It turns out that the dark matter induced antiproton flux already reaches or even exceeds the observation of AMS-02 at energy of $\sim 100$ GeV (see Figure~\ref{fig:excess-pbar}). The inclusion of the background resulting in the propagation of the high energy particles in the Galaxy can only render the tension much more seriously, as demonstrated in both Figure~\ref{fig:excess-pbar} and Figure~\ref{fig:contour}. In particular, the inferred annihilation cross section by~\citep{2025JCAP...11..080T} is about 40 times higher than the constraint set by the antiprotons after the proper inclusion of the astrophysical background (see Figure~\ref{fig:contour}). 
It is therefore quite reasonable to conclude that the observed 20 GeV $\gamma$-ray excess does not originate from dark matter annihilation.

Note that in previous searches for dark matter signal with the antiproton data, one serious challenge is the solar modulation effect 
that is hard to be accurately modeled~\citep{2025JCAP...10..049D}. However, for the current work, the dark matter mass is suggested to be $\sim 500$ GeV, hence the most prominent signal appears at $\sim 100$ GeV, for which the solar modulation effect is not strong any longer. In other words, our conclusion will not be changed by the adoption of other solar modulation models or the different approaches of the background calculation.



\section*{Acknowledgments}
We sincerely appreciate Professor Yi-Zhong Fan and Mr. Bo Zhang for helpful discussion.
This work is supported in part by the National Key Research and Development Program of China (No. 2022YFF0503304) and the Project for Young Scientists in Basic Research of the Chinese Academy of Sciences (No. YSBR-092).


\appendix


\bibliographystyle{JHEP}
\bibliography{biblio}

@ARTICLE{2009arXiv0910.2998G,
       author = {{Goodenough}, Lisa and {Hooper}, Dan},
        title = "{Possible Evidence For Dark Matter Annihilation In The Inner Milky Way From The Fermi Gamma Ray Space Telescope}",
      journal = {arXiv e-prints},
         year = 2009,
        month = oct,
          eid = {arXiv:0910.2998},
        pages = {arXiv:0910.2998},
          doi = {10.48550/arXiv.0910.2998},
archivePrefix = {arXiv},
       eprint = {0910.2998},
 primaryClass = {hep-ph},
       adsurl = {https://ui.adsabs.harvard.edu/abs/2009arXiv0910.2998G}
}

@ARTICLE{2008Natur.456..362C,
       author = {{Chang}, J. and {Adams}, J.~H. and {Ahn}, H.~S. and {Bashindzhagyan}, G.~L. and {Christl}, M. and {Ganel}, O. and {Guzik}, T.~G. and {Isbert}, J. and {Kim}, K.~C. and {Kuznetsov}, E.~N. and {Panasyuk}, M.~I. and {Panov}, A.~D. and {Schmidt}, W.~K.~H. and {Seo}, E.~S. and {Sokolskaya}, N.~V. and {Watts}, J.~W. and {Wefel}, J.~P. and {Wu}, J. and {Zatsepin}, V.~I.},
        title = "{An excess of cosmic ray electrons at energies of 300-800GeV}",
      journal = {\nat},
         year = 2008,
        month = nov,
       volume = {456},
       number = {7220},
        pages = {362-365},
          doi = {10.1038/nature07477},
       adsurl = {https://ui.adsabs.harvard.edu/abs/2008Natur.456..362C}
}

@ARTICLE{2009Natur.458..607A,
       author = {{Adriani}, O. and {Barbarino}, G.~C. and {Bazilevskaya}, G.~A. and {Bellotti}, R. and {Boezio}, M. and {Bogomolov}, E.~A. and {Bonechi}, L. and {Bongi}, M. and {Bonvicini}, V. and {Bottai}, S. and {Bruno}, A. and {Cafagna}, F. and {Campana}, D. and {Carlson}, P. and {Casolino}, M. and {Castellini}, G. and {de Pascale}, M.~P. and {de Rosa}, G. and {de Simone}, N. and {di Felice}, V. and {Galper}, A.~M. and {Grishantseva}, L. and {Hofverberg}, P. and {Koldashov}, S.~V. and {Krutkov}, S.~Y. and {Kvashnin}, A.~N. and {Leonov}, A. and {Malvezzi}, V. and {Marcelli}, L. and {Menn}, W. and {Mikhailov}, V.~V. and {Mocchiutti}, E. and {Orsi}, S. and {Osteria}, G. and {Papini}, P. and {Pearce}, M. and {Picozza}, P. and {Ricci}, M. and {Ricciarini}, S.~B. and {Simon}, M. and {Sparvoli}, R. and {Spillantini}, P. and {Stozhkov}, Y.~I. and {Vacchi}, A. and {Vannuccini}, E. and {Vasilyev}, G. and {Voronov}, S.~A. and {Yurkin}, Y.~T. and {Zampa}, G. and {Zampa}, N. and {Zverev}, V.~G.},
        title = "{An anomalous positron abundance in cosmic rays with energies 1.5-100GeV}",
      journal = {\nat},
         year = 2009,
        month = apr,
       volume = {458},
       number = {7238},
        pages = {607-609},
          doi = {10.1038/nature07942},
archivePrefix = {arXiv},
       eprint = {0810.4995},
 primaryClass = {astro-ph},
       adsurl = {https://ui.adsabs.harvard.edu/abs/2009Natur.458..607A}
}

@ARTICLE{2017Natur.552...63D,
       author = {{DAMPE Collaboration} and {Ambrosi}, G. and {An}, Q. and {Asfandiyarov}, R. and {Azzarello}, P. and {Bernardini}, P. and {Bertucci}, B. and {Cai}, M.~S. and {Chang}, J. and {Chen}, D.~Y. and {Chen}, H.~F. and {Chen}, J.~L. and {Chen}, W. and {Cui}, M.~Y. and {Cui}, T.~S. and {D'Amone}, A. and {de Benedittis}, A. and {De Mitri}, I. and {di Santo}, M. and {Dong}, J.~N. and {Dong}, T.~K. and {Dong}, Y.~F. and {Dong}, Z.~X. and {Donvito}, G. and {Droz}, D. and {Duan}, K.~K. and {Duan}, J.~L. and {Duranti}, M. and {D'Urso}, D. and {Fan}, R.~R. and {Fan}, Y.~Z. and {Fang}, F. and {Feng}, C.~Q. and {Feng}, L. and {Fusco}, P. and {Gallo}, V. and {Gan}, F.~J. and {Gao}, M. and {Gao}, S.~S. and {Gargano}, F. and {Garrappa}, S. and {Gong}, K. and {Gong}, Y.~Z. and {Guo}, D.~Y. and {Guo}, J.~H. and {Hu}, Y.~M. and {Huang}, G.~S. and {Huang}, Y.~Y. and {Ionica}, M. and {Jiang}, D. and {Jiang}, W. and {Jin}, X. and {Kong}, J. and {Lei}, S.~J. and {Li}, S. and {Li}, X. and {Li}, W.~L. and {Li}, Y. and {Liang}, Y.~F. and {Liang}, Y.~M. and {Liao}, N.~H. and {Liu}, H. and {Liu}, J. and {Liu}, S.~B. and {Liu}, W.~Q. and {Liu}, Y. and {Loparco}, F. and {Ma}, M. and {Ma}, P.~X. and {Ma}, S.~Y. and {Ma}, T. and {Ma}, X.~Q. and {Ma}, X.~Y. and {Marsella}, G. and {Mazziotta}, M.~N. and {Mo}, D. and {Niu}, X.~Y. and {Peng}, X.~Y. and {Peng}, W.~X. and {Qiao}, R. and {Rao}, J.~N. and {Salinas}, M.~M. and {Shang}, G.~Z. and {H. Shen}, W. and {Shen}, Z.~Q. and {Shen}, Z.~T. and {Song}, J.~X. and {Su}, H. and {Su}, M. and {Sun}, Z.~Y. and {Surdo}, A. and {Teng}, X.~J. and {Tian}, X.~B. and {Tykhonov}, A. and {Vagelli}, V. and {Vitillo}, S. and {Wang}, C. and {Wang}, H. and {Wang}, H.~Y. and {Wang}, J.~Z. and {Wang}, L.~G. and {Wang}, Q. and {Wang}, S. and {Wang}, X.~H. and {Wang}, X.~L. and {Wang}, Y.~F. and {Wang}, Y.~P. and {Wang}, Y.~Z. and {Wen}, S.~C. and {Wang}, Z.~M. and {Wei}, D.~M. and {Wei}, J.~J. and {Wei}, Y.~F. and {Wu}, D. and {Wu}, J. and {Wu}, L.~B. and {Wu}, S.~S. and {Wu}, X. and {Xi}, K. and {Xia}, Z.~Q. and {Xin}, Y.~L. and {Xu}, H.~T. and {Xu}, Z.~L. and {Xu}, Z.~Z. and {Xue}, G.~F. and {Yang}, H.~B. and {Yang}, P. and {Yang}, Y.~Q. and {Yang}, Z.~L. and {Yao}, H.~J. and {Yu}, Y.~H. and {Yuan}, Q. and {Yue}, C. and {Zang}, J.~J. and {Zhang}, C. and {Zhang}, D.~L. and {Zhang}, F. and {Zhang}, J.~B. and {Zhang}, J.~Y. and {Zhang}, J.~Z. and {Zhang}, L. and {Zhang}, P.~F. and {Zhang}, S.~X. and {Zhang}, W.~Z. and {Zhang}, Y. and {Zhang}, Y.~J. and {Zhang}, Y.~Q. and {Zhang}, Y.~L. and {Zhang}, Y.~P. and {Zhang}, Z. and {Zhang}, Z.~Y. and {Zhao}, H. and {Zhao}, H.~Y. and {Zhao}, X.~F. and {Zhou}, C.~Y. and {Zhou}, Y. and {Zhu}, X. and {Zhu}, Y. and {Zimmer}, S.},
        title = "{Direct detection of a break in the teraelectronvolt cosmic-ray spectrum of electrons and positrons}",
      journal = {\nat},
         year = 2017,
        month = dec,
       volume = {552},
       number = {7683},
        pages = {63-66},
          doi = {10.1038/nature24475},
archivePrefix = {arXiv},
       eprint = {1711.10981},
 primaryClass = {astro-ph.HE},
       adsurl = {https://ui.adsabs.harvard.edu/abs/2017Natur.552...63D}
}

@ARTICLE{2024arXiv240711737F,
       author = {{Fan}, Yi-Zhong and {Shen}, Zhao-Qiang and {Liang}, Yun-Feng and {Li}, Xiang and {Duan}, Kai-Kai and {Xia}, Zi-Qing and {Huang}, Xiao-Yuan and {Feng}, Lei and {Yuan}, Qiang},
        title = "{A $\sim 43$ GeV $\gamma$-ray line signature in the directions of a group of nearby massive galaxy clusters}",
      journal = {arXiv e-prints},
         year = 2024,
        month = jul,
          eid = {arXiv:2407.11737},
        pages = {arXiv:2407.11737},
          doi = {10.48550/arXiv.2407.11737},
archivePrefix = {arXiv},
       eprint = {2407.11737},
 primaryClass = {astro-ph.HE},
       adsurl = {https://ui.adsabs.harvard.edu/abs/2024arXiv240711737F}
}

@ARTICLE{2015PhRvD..91l3010Z,
       author = {{Zhou}, Bei and {Liang}, Yun-Feng and {Huang}, Xiaoyuan and {Li}, Xiang and {Fan}, Yi-Zhong and {Feng}, Lei and {Chang}, Jin},
        title = "{GeV excess in the Milky Way: The role of diffuse galactic gamma-ray emission templates}",
      journal = {\prd},
         year = 2015,
        month = jun,
       volume = {91},
       number = {12},
          eid = {123010},
        pages = {123010},
          doi = {10.1103/PhysRevD.91.123010},
archivePrefix = {arXiv},
       eprint = {1406.6948},
 primaryClass = {astro-ph.HE},
       adsurl = {https://ui.adsabs.harvard.edu/abs/2015PhRvD..91l3010Z}
}

@ARTICLE{2018RvMP...90d5002B,
       author = {{Bertone}, Gianfranco and {Hooper}, Dan},
        title = "{History of dark matter}",
      journal = {Reviews of Modern Physics},
         year = 2018,
        month = oct,
       volume = {90},
       number = {4},
          eid = {045002},
        pages = {045002},
          doi = {10.1103/RevModPhys.90.045002},
archivePrefix = {arXiv},
       eprint = {1605.04909},
 primaryClass = {astro-ph.CO},
       adsurl = {https://ui.adsabs.harvard.edu/abs/2018RvMP...90d5002B}
}

@ARTICLE{2021PrPNP.11903865A,
       author = {{Arbey}, A. and {Mahmoudi}, F.},
        title = "{Dark matter and the early Universe: A review}",
      journal = {Progress in Particle and Nuclear Physics},
         year = 2021,
        month = jul,
       volume = {119},
          eid = {103865},
        pages = {103865},
          doi = {10.1016/j.ppnp.2021.103865},
archivePrefix = {arXiv},
       eprint = {2104.11488},
 primaryClass = {hep-ph},
       adsurl = {https://ui.adsabs.harvard.edu/abs/2021PrPNP.11903865A}
}

@ARTICLE{2017ApJ...840...43A,
       author = {{Ackermann}, M. and {Ajello}, M. and {Albert}, A. and {Atwood}, W.~B. and {Baldini}, L. and {Ballet}, J. and {Barbiellini}, G. and {Bastieri}, D. and {Bellazzini}, R. and {Bissaldi}, E. and {Blandford}, R.~D. and {Bloom}, E.~D. and {Bonino}, R. and {Bottacini}, E. and {Brandt}, T.~J. and {Bregeon}, J. and {Bruel}, P. and {Buehler}, R. and {Burnett}, T.~H. and {Cameron}, R.~A. and {Caputo}, R. and {Caragiulo}, M. and {Caraveo}, P.~A. and {Cavazzuti}, E. and {Cecchi}, C. and {Charles}, E. and {Chekhtman}, A. and {Chiang}, J. and {Chiappo}, A. and {Chiaro}, G. and {Ciprini}, S. and {Conrad}, J. and {Costanza}, F. and {Cuoco}, A. and {Cutini}, S. and {D'Ammando}, F. and {de Palma}, F. and {Desiante}, R. and {Digel}, S.~W. and {Di Lalla}, N. and {Di Mauro}, M. and {Di Venere}, L. and {Drell}, P.~S. and {Favuzzi}, C. and {Fegan}, S.~J. and {Ferrara}, E.~C. and {Focke}, W.~B. and {Franckowiak}, A. and {Fukazawa}, Y. and {Funk}, S. and {Fusco}, P. and {Gargano}, F. and {Gasparrini}, D. and {Giglietto}, N. and {Giordano}, F. and {Giroletti}, M. and {Glanzman}, T. and {Gomez-Vargas}, G.~A. and {Green}, D. and {Grenier}, I.~A. and {Grove}, J.~E. and {Guillemot}, L. and {Guiriec}, S. and {Gustafsson}, M. and {Harding}, A.~K. and {Hays}, E. and {Hewitt}, J.~W. and {Horan}, D. and {Jogler}, T. and {Johnson}, A.~S. and {Kamae}, T. and {Kocevski}, D. and {Kuss}, M. and {La Mura}, G. and {Larsson}, S. and {Latronico}, L. and {Li}, J. and {Longo}, F. and {Loparco}, F. and {Lovellette}, M.~N. and {Lubrano}, P. and {Magill}, J.~D. and {Maldera}, S. and {Malyshev}, D. and {Manfreda}, A. and {Martin}, P. and {Mazziotta}, M.~N. and {Michelson}, P.~F. and {Mirabal}, N. and {Mitthumsiri}, W. and {Mizuno}, T. and {Moiseev}, A.~A. and {Monzani}, M.~E. and {Morselli}, A. and {Negro}, M. and {Nuss}, E. and {Ohsugi}, T. and {Orienti}, M. and {Orlando}, E. and {Ormes}, J.~F. and {Paneque}, D. and {Perkins}, J.~S. and {Persic}, M. and {Pesce-Rollins}, M. and {Piron}, F. and {Principe}, G. and {Rain{\`o}}, S. and {Rando}, R. and {Razzano}, M. and {Razzaque}, S. and {Reimer}, A. and {Reimer}, O. and {S{\'a}nchez-Conde}, M. and {Sgr{\`o}}, C. and {Simone}, D. and {Siskind}, E.~J. and {Spada}, F. and {Spandre}, G. and {Spinelli}, P. and {Suson}, D.~J. and {Tajima}, H. and {Tanaka}, K. and {Thayer}, J.~B. and {Tibaldo}, L. and {Torres}, D.~F. and {Troja}, E. and {Uchiyama}, Y. and {Vianello}, G. and {Wood}, K.~S. and {Wood}, M. and {Zaharijas}, G. and {Zimmer}, S. and {Fermi LAT Collaboration}},
        title = "{The Fermi Galactic Center GeV Excess and Implications for Dark Matter}",
      journal = {\apj},
         year = 2017,
        month = may,
       volume = {840},
       number = {1},
          eid = {43},
        pages = {43},
          doi = {10.3847/1538-4357/aa6cab},
archivePrefix = {arXiv},
       eprint = {1704.03910},
 primaryClass = {astro-ph.HE},
       adsurl = {https://ui.adsabs.harvard.edu/abs/2017ApJ...840...43A}
}

@ARTICLE{2015PhRvL.115w1301A,
       author = {{Ackermann}, M. and {Albert}, A. and {Anderson}, B. and {Atwood}, W.~B. and {Baldini}, L. and {Barbiellini}, G. and {Bastieri}, D. and {Bechtol}, K. and {Bellazzini}, R. and {Bissaldi}, E. and {Blandford}, R.~D. and {Bloom}, E.~D. and {Bonino}, R. and {Bottacini}, E. and {Brandt}, T.~J. and {Bregeon}, J. and {Bruel}, P. and {Buehler}, R. and {Caliandro}, G.~A. and {Cameron}, R.~A. and {Caputo}, R. and {Caragiulo}, M. and {Caraveo}, P.~A. and {Cecchi}, C. and {Charles}, E. and {Chekhtman}, A. and {Chiang}, J. and {Chiaro}, G. and {Ciprini}, S. and {Claus}, R. and {Cohen-Tanugi}, J. and {Conrad}, J. and {Cuoco}, A. and {Cutini}, S. and {D'Ammando}, F. and {de Angelis}, A. and {de Palma}, F. and {Desiante}, R. and {Digel}, S.~W. and {Di Venere}, L. and {Drell}, P.~S. and {Drlica-Wagner}, A. and {Essig}, R. and {Favuzzi}, C. and {Fegan}, S.~J. and {Ferrara}, E.~C. and {Focke}, W.~B. and {Franckowiak}, A. and {Fukazawa}, Y. and {Funk}, S. and {Fusco}, P. and {Gargano}, F. and {Gasparrini}, D. and {Giglietto}, N. and {Giordano}, F. and {Giroletti}, M. and {Glanzman}, T. and {Godfrey}, G. and {Gomez-Vargas}, G.~A. and {Grenier}, I.~A. and {Guiriec}, S. and {Gustafsson}, M. and {Hays}, E. and {Hewitt}, J.~W. and {Horan}, D. and {Jogler}, T. and {J{\'o}hannesson}, G. and {Kuss}, M. and {Larsson}, S. and {Latronico}, L. and {Li}, J. and {Li}, L. and {Llena Garde}, M. and {Longo}, F. and {Loparco}, F. and {Lubrano}, P. and {Malyshev}, D. and {Mayer}, M. and {Mazziotta}, M.~N. and {McEnery}, J.~E. and {Meyer}, M. and {Michelson}, P.~F. and {Mizuno}, T. and {Moiseev}, A.~A. and {Monzani}, M.~E. and {Morselli}, A. and {Murgia}, S. and {Nuss}, E. and {Ohsugi}, T. and {Orienti}, M. and {Orlando}, E. and {Ormes}, J.~F. and {Paneque}, D. and {Perkins}, J.~S. and {Pesce-Rollins}, M. and {Piron}, F. and {Pivato}, G. and {Porter}, T.~A. and {Rain{\`o}}, S. and {Rando}, R. and {Razzano}, M. and {Reimer}, A. and {Reimer}, O. and {Ritz}, S. and {S{\'a}nchez-Conde}, M. and {Schulz}, A. and {Sehgal}, N. and {Sgr{\`o}}, C. and {Siskind}, E.~J. and {Spada}, F. and {Spandre}, G. and {Spinelli}, P. and {Strigari}, L. and {Tajima}, H. and {Takahashi}, H. and {Thayer}, J.~B. and {Tibaldo}, L. and {Torres}, D.~F. and {Troja}, E. and {Vianello}, G. and {Werner}, M. and {Winer}, B.~L. and {Wood}, K.~S. and {Wood}, M. and {Zaharijas}, G. and {Zimmer}, S. and {Fermi-LAT Collaboration}},
        title = "{Searching for Dark Matter Annihilation from Milky Way Dwarf Spheroidal Galaxies with Six Years of Fermi Large Area Telescope Data}",
      journal = {\prl},
         year = 2015,
        month = dec,
       volume = {115},
       number = {23},
          eid = {231301},
        pages = {231301},
          doi = {10.1103/PhysRevLett.115.231301},
archivePrefix = {arXiv},
       eprint = {1503.02641},
 primaryClass = {astro-ph.HE},
       adsurl = {https://ui.adsabs.harvard.edu/abs/2015PhRvL.115w1301A}
}

@ARTICLE{2017ApJ...834..110A,
       author = {{Albert}, A. and {Anderson}, B. and {Bechtol}, K. and {Drlica-Wagner}, A. and {Meyer}, M. and {S{\'a}nchez-Conde}, M. and {Strigari}, L. and {Wood}, M. and {Abbott}, T.~M.~C. and {Abdalla}, F.~B. and {Benoit-L{\'e}vy}, A. and {Bernstein}, G.~M. and {Bernstein}, R.~A. and {Bertin}, E. and {Brooks}, D. and {Burke}, D.~L. and {Carnero Rosell}, A. and {Carrasco Kind}, M. and {Carretero}, J. and {Crocce}, M. and {Cunha}, C.~E. and {D'Andrea}, C.~B. and {da Costa}, L.~N. and {Desai}, S. and {Diehl}, H.~T. and {Dietrich}, J.~P. and {Doel}, P. and {Eifler}, T.~F. and {Evrard}, A.~E. and {Fausti Neto}, A. and {Finley}, D.~A. and {Flaugher}, B. and {Fosalba}, P. and {Frieman}, J. and {Gerdes}, D.~W. and {Goldstein}, D.~A. and {Gruen}, D. and {Gruendl}, R.~A. and {Honscheid}, K. and {James}, D.~J. and {Kent}, S. and {Kuehn}, K. and {Kuropatkin}, N. and {Lahav}, O. and {Li}, T.~S. and {Maia}, M.~A.~G. and {March}, M. and {Marshall}, J.~L. and {Martini}, P. and {Miller}, C.~J. and {Miquel}, R. and {Neilsen}, E. and {Nord}, B. and {Ogando}, R. and {Plazas}, A.~A. and {Reil}, K. and {Romer}, A.~K. and {Rykoff}, E.~S. and {Sanchez}, E. and {Santiago}, B. and {Schubnell}, M. and {Sevilla-Noarbe}, I. and {Smith}, R.~C. and {Soares-Santos}, M. and {Sobreira}, F. and {Suchyta}, E. and {Swanson}, M.~E.~C. and {Tarle}, G. and {Vikram}, V. and {Walker}, A.~R. and {Wechsler}, R.~H. and {Fermi-LAT Collaboration} and {DES Collaboration}},
        title = "{Searching for Dark Matter Annihilation in Recently Discovered Milky Way Satellites with Fermi-Lat}",
      journal = {\apj},
         year = 2017,
        month = jan,
       volume = {834},
       number = {2},
          eid = {110},
        pages = {110},
          doi = {10.3847/1538-4357/834/2/110},
archivePrefix = {arXiv},
       eprint = {1611.03184},
 primaryClass = {astro-ph.HE},
       adsurl = {https://ui.adsabs.harvard.edu/abs/2017ApJ...834..110A}
}

@ARTICLE{2020ARNPS..70..455M,
       author = {{Murgia}, Simona},
        title = "{The Fermi{\textendash}LAT Galactic Center Excess: Evidence of Annihilating Dark Matter?}",
      journal = {Annual Review of Nuclear and Particle Science},
         year = 2020,
        month = oct,
       volume = {70},
        pages = {455-483},
          doi = {10.1146/annurev-nucl-101916-123029},
       adsurl = {https://ui.adsabs.harvard.edu/abs/2020ARNPS..70..455M}
}

@ARTICLE{2025JCAP...11..080T,
       author = {{Totani}, Tomonori},
        title = "{20 GeV halo-like excess of the Galactic diffuse emission and implications for dark matter annihilation}",
      journal = {\jcap},
         year = 2025,
        month = nov,
       volume = {2025},
       number = {11},
          eid = {080},
        pages = {080},
          doi = {10.1088/1475-7516/2025/11/080},
archivePrefix = {arXiv},
       eprint = {2507.07209},
 primaryClass = {astro-ph.HE},
       adsurl = {https://ui.adsabs.harvard.edu/abs/2025JCAP...11..080T}
}

@ARTICLE{2025arXiv251201404M,
       author = {{Murayama}, Hitoshi},
        title = "{Resonant Annihilation of WIMP Dark Matter for Halo Gamma Ray Signal}",
      journal = {arXiv e-prints},
         year = 2025,
        month = dec,
          eid = {arXiv:2512.01404},
        pages = {arXiv:2512.01404},
          doi = {10.48550/arXiv.2512.01404},
archivePrefix = {arXiv},
       eprint = {2512.01404},
 primaryClass = {hep-ph},
       adsurl = {https://ui.adsabs.harvard.edu/abs/2025arXiv251201404M}
}

@ARTICLE{2017PhRvL.118s1101C,
       author = {{Cui}, Ming-Yang and {Yuan}, Qiang and {Tsai}, Yue-Lin Sming and {Fan}, Yi-Zhong},
        title = "{Possible Dark Matter Annihilation Signal in the AMS-02 Antiproton Data}",
      journal = {Physical Review Letters},
         year = 2017,
        month = may,
       volume = {118},
       number = {19},
          eid = {191101},
        pages = {191101},
          doi = {10.1103/PhysRevLett.118.191101},
archivePrefix = {arXiv},
       eprint = {1610.03840},
 primaryClass = {astro-ph.HE},
       adsurl = {https://ui.adsabs.harvard.edu/abs/2017PhRvL.118s1101C}
}

@ARTICLE{2017PhRvL.118s1102C,
       author = {{Cuoco}, Alessandro and {Kr{\"a}mer}, Michael and {Korsmeier}, Michael},
        title = "{Novel Dark Matter Constraints from Antiprotons in Light of AMS-02}",
      journal = {Physical Review Letters},
         year = 2017,
        month = may,
       volume = {118},
       number = {19},
          eid = {191102},
        pages = {191102},
          doi = {10.1103/PhysRevLett.118.191102},
archivePrefix = {arXiv},
       eprint = {1610.03071},
 primaryClass = {astro-ph.HE},
       adsurl = {https://ui.adsabs.harvard.edu/abs/2017PhRvL.118s1102C}
}

@ARTICLE{2022PhRvL.129w1101Z,
       author = {{Zhu}, Cheng-Rui and {Cui}, Ming-Yang and {Xia}, Zi-Qing and {Yu}, Zhao-Huan and {Huang}, Xiaoyuan and {Yuan}, Qiang and {Fan}, Yi-Zhong},
        title = "{Explaining the GeV Antiproton Excess, GeV {\ensuremath{\gamma}} -Ray Excess, and W -Boson Mass Anomaly in an Inert Two Higgs Doublet Model}",
      journal = {Physical Review Letters},
         year = 2022,
        month = dec,
       volume = {129},
       number = {23},
          eid = {231101},
        pages = {231101},
          doi = {10.1103/PhysRevLett.129.231101},
archivePrefix = {arXiv},
       eprint = {2204.03767},
 primaryClass = {astro-ph.HE},
       adsurl = {https://ui.adsabs.harvard.edu/abs/2022PhRvL.129w1101Z}
}

@ARTICLE{2025JCAP...10..049D,
       author = {{Duan}, Kai-Kai and {Wang}, Xiao and {Li}, Wen-Hao and {Xu}, Zhi-Hui and {Sming Tsai}, Yue-Lin and {Fan}, Yi-Zhong},
        title = "{Scrutinizing the impact of the solar modulation on AMS-02 antiproton excess}",
      journal = {\jcap},
         year = 2025,
        month = oct,
       volume = {2025},
       number = {10},
          eid = {049},
        pages = {049},
          doi = {10.1088/1475-7516/2025/10/049},
archivePrefix = {arXiv},
       eprint = {2506.13352},
 primaryClass = {astro-ph.CO},
       adsurl = {https://ui.adsabs.harvard.edu/abs/2025JCAP...10..049D}
}

@ARTICLE{1968ApJ...154.1011G,
       author = {{Gleeson}, L.~J. and {Axford}, W.~I.},
        title = "{Solar Modulation of Galactic Cosmic Rays}",
      journal = {Astrophysical Journal},
         year = 1968,
        month = dec,
       volume = {154},
        pages = {1011},
          doi = {10.1086/149822},
       adsurl = {https://ui.adsabs.harvard.edu/abs/1968ApJ...154.1011G}
}
\end{document}